\documentclass{egpubl}

 \JournalSubmission    
 \electronicVersion 

\ifpdf \usepackage[pdftex]{graphicx} \pdfcompresslevel=9
\else \usepackage[dvips]{graphicx} \fi

\PrintedOrElectronic
\usepackage{algorithm}
\usepackage{algorithmic}
\usepackage{t1enc,dfadobe}

\usepackage{egweblnk}
\usepackage{cite}
\usepackage{amsmath}
\usepackage{graphicx}
\usepackage{parskip}
\usepackage{epstopdf}
        \let\tt=\ttfamily
\let\it=\itshape     \let\sl=\slshape     
\let\bf=\bfseries

\title{Interactive Character Posing by Sparse Coding}
\author[Ranch Y.Q. Lai \& Pong C. Yuen \& Kelvin K.W. Lee \& JH Lai]
       {Ranch Y.Q. Lai $^{1}$
       and Pong C. Yuen $^{1}$
        and Kelvin K.W. Lee$^{2}$
        and JH Lai$^{3}$
        \\
         $^1$Department of Computer Science, Hong Kong Baptist University, Hong Kong, China\\
         $^2$Department of Communication studies, Hong Kong Baptist University, Hong Kong, China\\
         $^3$School of Information Science and Technology, Sun Yat-sen University, GuangZhou, China         }
      
\begin{document}

\maketitle

\begin{abstract}

Character posing is of interest in computer animation.  It is difficult due to its dependence on  inverse kinematics (IK) techniques and articulate property of human characters .  To solve the IK problem, classical methods that rely on numerical solutions often suffer from the under-determination problem and can not guarantee naturalness. Existing data-driven methods address this problem by learning from motion capture data. When facing a large variety of poses however,  these methods may not be able to capture the pose styles or be applicable in real-time environment.  
Inspired from the low-rank motion de-noising and completion model in \cite{lai2011motion}, we propose a novel model for character posing based on sparse coding. Unlike conventional approaches,  our model directly captures the pose styles in Euclidean space to provide intuitive training error measurements and facilitate pose synthesis. A pose dictionary is learned in training stage and based on it natural poses are synthesized to satisfy users' constraints . 
We compare our model with existing models for tasks of pose de-noising and completion.  Experiments show our model obtains lower de-noising and completion error. We also provide User Interface(UI) examples illustrating that our model is effective for interactive character posing.

\begin{classification} 
\CCScat{Computer Graphics}{I.7}{Computer Graphics}{Animation}
\end{classification}
\end{abstract}

\section{Introduction}

Character posing is an important step for key-frame animation. It is difficult for novices and even skilled artists due to the articulate property of human motion. With the most prevailing input device still being the mouse,    the users' input can only provide basic information such as 2D screen coordinates.  Based on this limited information,   it is challenging to generates satisfactory character poses efficiently.  Moreover,   it requires considerable information to determine the character's all degrees of freedom (DOF's). Given some 3D positional information of one or some joints,   it is useful to re-position the rest of joints or even the whole pose if the information provided by users is not accurate. For example, a novice animator is likely to pose an unnatural character within a short time limit. It is then up to the algorithm to extract useful information such as pose style from the unnatural character and create a new natural one. 
This shall carry out interactively for synthesizing natural poses, and the process is referred to as character posing. 

To solve the character posing problem, inverse kinematics is often necessary to find the skeleton in angle space representation. The classical inverse kinematics solves an under-determined non-linear system to find  the joint angles. One popular method is to exploit the gradient information--that is,   to construct the Jacobian matrix and then solve the system iteratively starting from a random initial point. However,   the mapping from  3D Euclidean space to joint angle space is one-to-many if the users' constraints are insufficient. For example, given a set of \textit{incomplete} joint constraints such as   the 3D positions of some joints,  the solution obtained from Jacobian method will not be unique but depends on the initiation,   not to mention that all poses resulting from the possible solutions are unlikely to be natural. One not only has to narrow down the solution set,   but must also refine the solutions so that the resulting pose is natural.

One way that may help is by learning from motion capture data.   Even though the space of joint configuration is large,  the desirable poses only span a much smaller space. For example,   human beings can make a large number poses,   but the space of natural  poses is smaller.  By recording  these poses in motion capture data and learning from them,   we can provide heuristics for solving the IK problem. This is the recent approach taken by researchers and is referred to as data-driven IK. Our approach falls in this category.

\begin{figure}
\centering
\includegraphics[width=\linewidth]{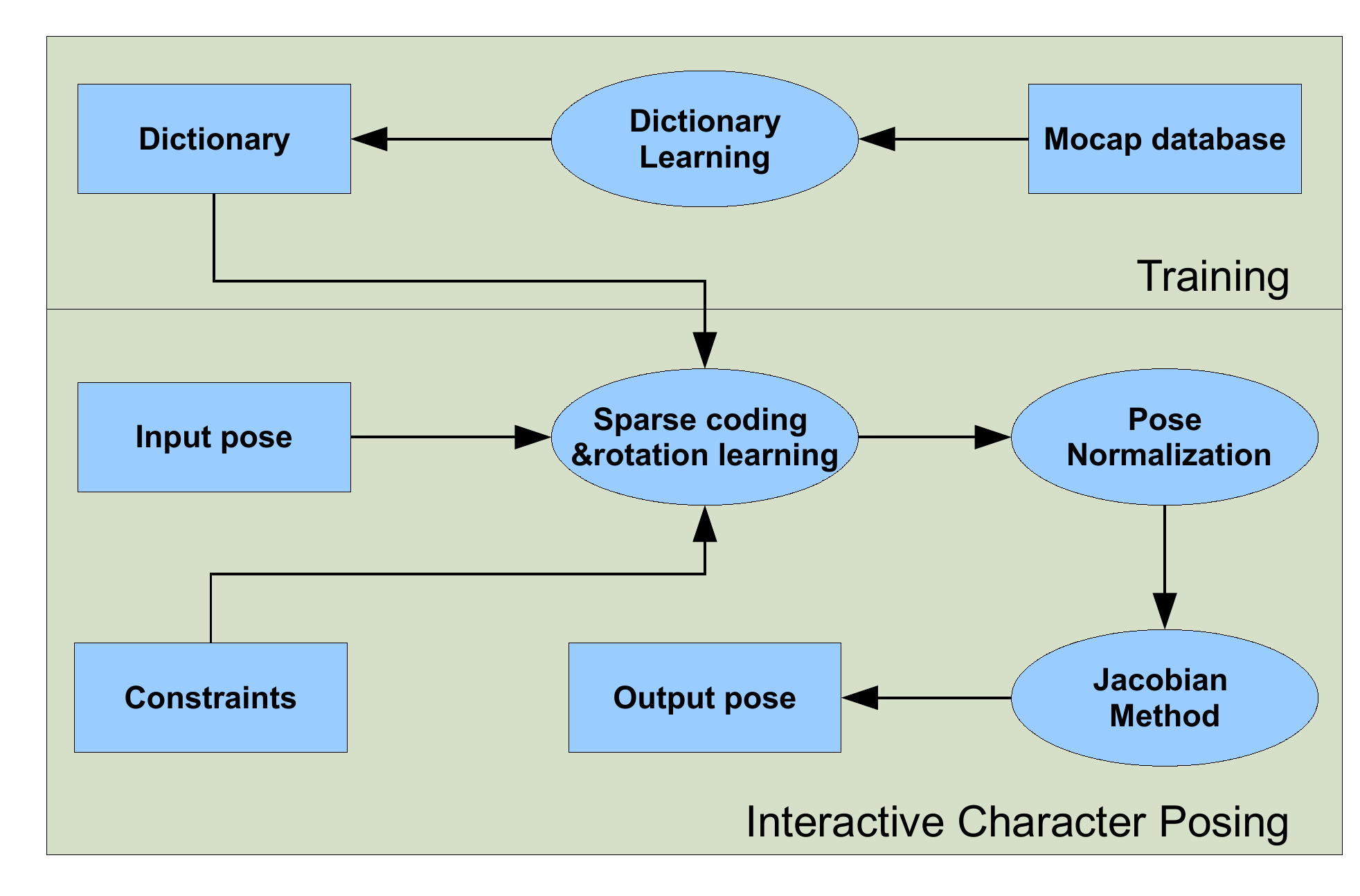}
\caption{Framework of proposed approach}
\label{fig_framework}
\end{figure}

The framework of our model is showed in figure \ref{fig_framework}. In our proposed model, each pose is assumed to have sparse representation given a pose dictionary. The pose dictionary is learned from motion capture data in Euclidean space. These data cover a large number of different motion styles,   such as walking,   running and other sports activities.  In the interactive character posing stage,   our model can  respond to the users' inputs and constraints in real-time and construct natural poses that meet the users' intentions.    
We solve the pose synthesis problem by  breaking the optimization problem into three components: 1) finding the sparse coefficient and rotation parameters; 2) normalizing the pose to determine the scaling parameter and 3) building the output pose in angle space by Jacobian method. Details on our model and how to solve the optimization problem are presented in section 3.


The rest of this paper is organized as follows. We review the related work in the remaining of this section. Starting from the subspace models, we present the inspiration for this work and derive our model in section 2.  In section 3 we introduce our proposed model and the algorithm for solving the model in detail,   followed by applications and experimental results in section 4. We present some discussions and conclude our work in section 5. 

\subsection{Related work}

\textbf{Classical inverse kinematics} The use of Jacobian matrix for inverse kinematics can be at least traced back to \cite{girard1985computational},   in which Girard et al linearised the
equation $\mathbf{p} = f (\mathbf{q})$ at current estimate $\mathbf{q}$,   yielding $\mathbf{p} = f (\mathbf{q})+J (\mathbf{q})^T\Delta\mathbf{q}$,   where $f$ is the forward-kinematic function which involves a set of translations and rotations, usually implemented procedurally in some programming language; $J (\mathbf{q})$ is the Jacobian matrix defined as $J (\mathbf{q}) = \frac{\partial f (\mathbf{q})}{\partial \mathbf{q}}$ . The Jacobian matrix  is usually not  full-rank and the update $\Delta\mathbf{q}$ is    given by $ 
\Delta\mathbf{q} = J (\mathbf{q})^\dagger  (\mathbf{p} - f (\mathbf{q}))$, 
where $J (\mathbf{q})^\dagger =  (J (\mathbf{q})J (\mathbf{q})^T + \eta \mathbf{I})^{-1}J (\mathbf{q})$ is the pseudo-inverse of the Jacobian matrix and $\eta$ is a small positive number. 
To accommodate constraints such as angle limit or spatial relations ,   Zhao et al \cite{zhao1994inverse} minimized the $\ell_2$ distance between input pose and forward-kinematics function subject to the constraints using non-linear programming technique. Rose et al\cite{rose1997process}  solved the IK problem using  BFGS optimization method \cite{gill1981practical}, which is an quasi-Newton algorithm that does not require the complicated Hessian matrix. Their work aimed at building      the final motion in angle representation using the sensor data,   which were obtained from motion capture  process in Euclidean space. 

%


\textbf{Data-driven Inverse Kinematics} In general,   data-driven IK leverages  the mocap data and  models the IK problems as follows
\begin{eqnarray}
\min. &E_{\text{Prior}} + \lambda E_{\text{IK}} \label{eqn_general_data_driven}
\end{eqnarray}

where $E_{\text{Prior}}$ is the energy term that measures the  (negative log) observation likelihood. $E_{\text{Prior}}$ is the energy term that measures the  (negative log) probability of current pose under some prior distribution. This prior distribution is what makes the model distinctive from other ones.

A straightforward model for modelling the prior is to use the Gaussian distribution.  Due to the connection in the covariance matrix,  this approach is related to Principal Component analysis (PCA) which restricts the solution to lie in the subspace span by the  principal components. By imposing the Gaussian prior, we force the solution to approach the mean from the direction of one principal component or a linear combination of them. Instead of using the Gaussian model directly, we can also first partition the motion data by clustering algorithm and then build a Gaussian prior for each cluster. This is similar to the mixture of local linear models which have been used as baseline models in  \cite{chai2005performance}.

Wei et al \cite{wei2011intuitive} modelled the prior using the mixture of factor analyzers  (MFA)  \cite{ghahramani1996algorithm}. MFA is similar to Mixture of Gaussian,   but includes a dimension reduction component and avoids the ill-condition problem of covariance matrix.  Kallmann et al  \cite{kallmann2008analytical} introduced analytical IK for the arms instead of the whole body. To construct a natural whole-body,   a set of pre-designed key body poses are used for pose blending (interpolation). 
Grochow et al \cite{grochow2004style} proposed style-based IK  (SIK) for modelling human motion.  Their model is based on the scaled Gaussian Processes Latent variable Models (GPLVM)  \cite{lawrence2004gaussian}. Specifically,   the training samples and the target poses are mapped to low-dimensional latent variables using GPLVM. These variables are connected by a kernel function. The information then passes from the training set through the correlation of latent variables to the target pose. Their model can generalize to unseen poses thanks to the good generalization ability of Gaussian processes. However,   due to the limitation of Gaussian processes,    the complexity of their approach is asymptotically  cubic to the size of training set. To reduce complexity, they maintained an active set during the training and synthesis. Despite the improved efficiency brought by the active set, it is still prohibitive to learn from large scale pose data for real-time applications.  To further improve efficiency, Wu et al \cite{wu2011natural} considered using different approximations to the speed up Gaussian processes and apply them to solve the IK problem.  Other than GPLVM and its variants,   modelling the motion based on dimension reduction is also very popular. Examples are those methods that based on state model \cite{brand2000style,  li2002motion} and PCA \cite{safonova2004synthesizing},   etc.

Given incomplete measurements of motion capture sensors,   Chai et al \cite{chai2005performance} construct a full-body human motion use Local PCA (LPCA) model. This basic idea is to incrementally estimate the current pose based on the previous estimated poses and a motion capture database. 
 This prior term  measures the deviation of reconstructed poses from the motion capture database.  Since the database can be large and heterogeneous,    they introduce a LPCA model: given an incomplete pose,   they first search the database to find the nearest pose and build a Gaussian motion prior around the neighbourhood.  The prior is then used for pose synthesis. 
 
Motion data de-noising and completion was considered by Lou et al \cite{lou2010example}. The idea is to first construct a set of filter bases from the motion capture data and use them for motion completion or de-noising. The resulting motion is the solution to a cost function that consists of the  bases-representation error and the observation likelihood. The filter bases capture spatial-temporal patterns of the human motion,   and the den-noising process also relies on the spatial-temporal patterns,  which in our case do not exist. Another work on this direction was introduced by Lai et al \cite{lai2011motion},   in which low-rank matrix completion algorithm was used for unsupervised motion de-noising and completion. The major difference of our working environment from both of these is that we do not have temporal information available when synthesizing new poses.

%
%
%
%
%
%
%


\textbf{Sparse coding} On the other side,   sparse representation has been widely applied to image processing and pattern recognition. Examples include face recognition \cite{wright2009robust}, image super-resolution \cite{yang2008image}, etc. For modelling human motion,   \cite{li2010learning} considered each joint's movement as a signal that admits sparse representation over a set of basis functions. These basis functions are learned from the motion capture data. They demonstrated that the proposed model is useful for action retrieval and classification.  Our work is different as we model each pose separately and our target application is on character posing instead of action retrieval and classification.

\textbf{Summary} Among the existing models, the numerical IK algorithms do not have access to motion capture data,   thus can not guarantee the naturalness of synthesized poses. For data-driven models,   Gaussian model will introduce large error for large dataset as it tries to approximate the underlying complicated distribution by Gaussian. Imposing a clustering step before applying Gaussian is an improvement but leaves us the problem of choosing the cluster number. For LPCA,  searching the pose in on-line mode is too slow for learning from large training set. The model accuracy also depends on both the searching result and the neighbourhood size.  For SIK,   the complexity problem is prohibitive for moderate-scale training set,   and the introduced active-set approximation is difficult to capture the diversity of motion styles. For MFA,   the introduction of a diagonal matrix in the covariance avoids the ill-condition problem when the cluster number is large (and thus the pose number in each cluster is small). However, being a variant of Gaussian model, it is still at the risk of under-fitting the complicated data. Apart from the limitations mentioned above, most of these models are probabilistic and the training error is measured by likelihood,   which is not intuitive: given such a measurement,   it is not straight-forward to determine whether the model is adequate for fitting the data. Consequently,   it is hard to choose the model parameters (e. g. ,   the cluster number). In contrast,   our model measures the training error directly by the mean square error,   which is very intuitive for determining model parameters. Besides,   our model can learn from large datasets (up to millions of poses) with an arbitrarily small training error (although) at the expense of the increasing the size of dictionary. We found that this increase does not cause the over-fitting problem and the de-noising and completion algorithm still maintains efficiency, as the complexity of our synthesis algorithm is linear to the size of pose dictionary .  We present some real-time applications demonstrating that our model is effective for interactive character posing.  We also compare our model with the existing models to test the performance of pose completion when a large proportion of joints are missing,   and pose de-noising when the pose is corrupted by dense and sparse Gaussian noise. Experimental results show that our model has lower completion and de-noising error.

\textbf{Contributions} Our contributions are two-fold: 1) starting from the prevailing subspace models in modelling human motion, we propose  sparse representation of poses for character posing; to our knowledge, we are the first to propose sparse coding for character posing; 2)       different from previous approaches, we propose to learn from the motion capture data in Euclidean space, which not only provides intuitive measurements in training error, but facilitate sparse coding and pose synthesis.

\section{Overview of proposed model}
\textbf{Notation setting}
In this paper,   matrices,  vectors and scalers are denoted in bold face upper-case,   bold face lower-case and non-bold lower-case letters respectively.  $||\cdot||_2$, $||\cdot||_0$ and $||\cdot||_F$ denote the vector $\ell_2$-norm, vector $\ell_0$-(pseudo)norm and matrix Frobenius norm respectively. $||\cdot||_0$  measures the number of non-zero components in a vector.

\textbf{From low-rank approximation of motion to sparse representation of poses}
As the movements of the body parts are correlated,   when we represent the human motion as a matrix,  it will be approximately low-rank,   endowing a fast-decaying spectrum\cite{lai2011motion}.  Low-rank approximation is therefore effective for modelling human motion. 
Our work on this paper is inspired from the low-rank motion  completion approach proposed by Lai et al\cite{lai2011motion}, in which the rank of a motion is minimized for completing and de-noising human motion. The connection between rank function and the $\ell_0$-norm minimization is clear if we observe that the rank of a diagonal matrix is equivalent to the $\ell_0$-norm of the diagonal vector.

Let $\{\mathbf{y}_i \in \mathcal{R}^D\}_{i=1,. . . ,m}$ denote a set of poses and $\mathbf{Y}$ be the motion matrix: $\mathbf{Y} = [\mathbf{y}_1,. . . ,\mathbf{y}_m]$. The Singular Value Decomposition (SVD) of $\mathbf{Y}$ gives $\mathbf{Y} = [\mathbf{u}_1,. . \mathbf{u}_D]\text{diag} ([s_1,. . . ,s_D])[\mathbf{v}_1,. . \mathbf{v}_D]^T$, where $\mathbf{u}_i \in \mathcal{R}^D, \mathbf{v}_i\in \mathcal{R}^m$. By setting $s_{k+1}. . . s_{D}$ to zeros, we can approximate each pose $\mathbf{y}_i$ by the first $k$ singular vectors:  $\mathbf{\hat{y}}_i = \sum_{j=1}^{k}\mathbf{u}_j\alpha_j^{(i)}$, where $\alpha_j^{(i)} = s_j\mathbf{v}_j^{ (i)}$ and  $\mathbf{v}_j^{ (i)}$ is the $i$th component of $\mathbf{v}_j$. The number $k$ can be small and we still have a good approximation for human motions.  Another words, each pose has a sparse representation given the set of bases $\{\mathbf{u}_j\}_{j=1,. . . , D}$, and the supports in this case lie in the first $k$ bases.

Following this idea, to learn from a large motion capture dataset, one possible way is to first partition the whole training set into $K$ clusters and find a set of bases $\mathbf{U}_i$ for cluster $i$ . If we then collect all the bases into a matrix, i.e.,  $\mathbf{U}=[\mathbf{U}_1,...,\mathbf{U}_K]$,  each pose in the training set will be  sparsely represented under such a matrix. This approach is general, as we can set $K$ to $1$ to get back to original low-rank approximation of the whole dataset, and set $K$ to the size of the dataset to use the whole dataset as bases. However, this leaves us the problem to find a way to properly partition the data into clusters. If the poses in a cluster are too linearly-uncorrelated because of improper partition(e.g., too many diversified poses in a cluster), then the sparse approximation error will tend to be large. On the other hand, if each cluster only consists of a few poses , the sparse approximation is small but the number  of  bases will be very large, with the limit being the size of training set.

Instead of determining the matrix $\mathbf{U}$ in the above means,  we take another way around: we learn the matrix from the training set without being worried about the partitioning. To begin with, we return to one-cluster case and note that from an optimization perspective, the bases $\mathbf{U} = [\mathbf{u}_1,...,\mathbf{u}_k]$ obtained from SVD is a solution to the following optimization problem  with variables $\mathbf{U}\in \mathcal{R}^{D\times k}$ and $\mathbf{X} \in \mathcal{R}^{k \times m}$
\begin{eqnarray}
&\min. & ||\mathbf{Y} - \mathbf{U}\mathbf{X}||_F^2 \label{eqn_svd}\\
& \text{s.t.}&\mathbf{U}^T\mathbf{U} = \mathbf{I}\nonumber
\end{eqnarray}
As a relaxation, the orthonormal constraint on $\mathbf{U}$ is changed to unit ball constraint on its columns and the size of $\mathbf{U}$ is extended to be $D \times n$ pose dictionary with $n > D$. However, each column of $\mathbf{X}$  shall be sparse to reflect the sparse property of poses. The sparsity constraint is measured by the $\ell_0$-norm of each column of $\mathbf{X}$. We refer to the matrix $\mathbf{U}$ in this case as pose dictionary and denote it as $
\mathbf{A}$ to be consistent with the sparse coding literature. We present this pose dictionary learning process in next section.

\textbf{Modelling the poses in Euclidean space} In the analysis presented above, we have no assumption on the ambient space  of poses. Although the motion matrix is low-rank in both Euclidean space and angle space (when preprocessed properly) ,    we choose the former for sparse representation. This is different from previous data-driven approaches,   which directly model the motion in angle space. We do so for two reasons.

One reason is  to avoid the periodicity of angles,   which  potentially corrupts the sparse representation:  given two identical poses and add $2\pi$ to the one pose vector while leaving the other the same,   then the resulting two poses are still identical,   but the $2\pi$-shifted one is unlikely to have same sparse representation as the other under the same dictionary.  This will be a problem especially in pose synthesis, due to the non-smoothness of $\ell_0$-norm.  The same will not happen for poses represented in Euclidean space. Other parametrizations such as quaternion and exponential map are  (also) non-linear, and thus inconvenient for sparse coding which involves solving a linear system. 

The other reason is that by doing so, we can directly measure the representation error, which provides us an intuitive measurements in training. Specifically, we are optimizing directly the (mean) square error of the sparse representation of the training set without invoking the forward-kinematics mapping. Moreover, since the input observations such as an edited poses and  2D/3D coordinates are in Euclidean space, the optimization for pose synthesis will be more efficient because we can defer the demand of Jacobian matrix till we find a pose represented by a full set of joint coordinates. And the need for converting the pose into angle space in the last stage( see figure \ref{fig_framework}) is only necessary when we want to further process the pose such as changing the skeleton configuration(e.g. joint angle limits, bone length).

\section{Learning sparse representation of poses for character posing}

The idea of modelling the poses based on sparse coding is similar to the subspace approaches such as PCA, except that the 'subspace' is generalized to the span of active atoms in the pose dictionary and the 'bases' are no longer assumed to be orthonormal or even independent.  Given a set of observations and constraints, the reconstructed pose shall be a  trade-off between having a sparse supports under the pose dictionary and being consistent with the observations and constraints.

\subsection{Learning the pose dictionary}
Before applying sparse representation,   we need to first determine the underlying dictionary, which should be able to capture the pose variations and be insensitive to global orientation and translation. 
 Given a training set  $\{\mathbf{y}_i\in \mathcal{R}^D\}_{i=1,. . . m}$,  the pose dictionary $\mathbf{A}\in \mathcal{R}^{D \times n}$ is learned such that the poses in the training set are sparsely represented under this dictionary. Specifically,   the learning problem is modelled as
\begin{eqnarray}
&\min. & ||\bar{\mathbf{Y}} - \mathbf{A}\mathbf{X}||_F^2 \label{eqn_dict_obj}\\
&\text{s. t. }& ||\mathbf{x}_j||_0 < \kappa,   j=1,  . . . ,  m\nonumber\\
&&||\mathbf{a}_k||_2 = 1,   k=1,  . . . ,  n\nonumber
\end{eqnarray}
where $\mathbf{X} = [\mathbf{x}_1,. . . ,\mathbf{x}_m], \bar{\mathbf{Y}} = [\bar{\mathbf{y}}_1,. . . ,\bar{\mathbf{y}}_m]$ and $\bar{\mathbf{y}}_i$ is the $i$th pose in training set,   with global orientation and translation set to zeros since they are usually irrelevant in affecting the pose style.  Note the similarity between problem \eqref{eqn_dict_obj} and  \eqref{eqn_svd}.

To solve the above learning problem,   we use the K-SVD algorithm proposed by Aharon et al \cite{aharon2006ksvd}.  K-SVD alternates between sparse coding and dictionary updating in every iterate. Specifically,   in the sparse coding stage,   the Orthogonal matching pursuit (OMP) \cite{pati1993orthogonal} is used to find a sparse representation of the training set,   while in the dictionary updating stage,   the columns  of the dictionary are updated sequentially by computing the singular value decomposition of the sparse coding residual matrix. It is reported that this method is better than the naive method of simply computing the least square solution by fixing $\mathbf{X}$ to update $\mathbf{A}$. We also refer readers to  \cite{aharon2006ksvd} for details.

\subsection{Pose synthesis problem}
Now by assuming that the pose dictionary $\mathbf{A}$ is given, we propose the following model for pose reconstruction: 
\begin{eqnarray}
 (P_0)&\min. &\gamma||\mathbf{Ax}-s\cdot f (\mathbf{q})||_2^2+||\mathbf{P}\{\tau\circ (s\cdot f (\mathbf{q})) - \bar{\mathbf{y}}_0\}||_2^2\nonumber\\
&&+||\mathbf{W} (\tau-\tau_0)||_2^2 \label{eqn_overall_objective} \\
&\text{s. t. }& ||\mathbf{x}||_0 \le \kappa \label{sp_constraint}\\
&& s>0
\end{eqnarray}

In the objective \eqref{eqn_overall_objective}, the first term measures the difference between sparse representation and the  forward-kinematics function $f$ scaled by a positive factor $s$. 
The scale $s$ is applied to all 3 dimensions in Euclidean space to maintain the skeleton scaling ratio. The constraint \eqref{sp_constraint} guarantees the sparsity of $\mathbf{x}$ in the solution.   

The second term measures the sparse coding error of the input under a rigid-body rotation. $\tau \in \mathcal{R}^3$ is the rotation parameter. The notation $\tau \circ (\mathbf{t}) $ denotes a 3D rotation of the vector $\mathbf{t}$ which is  concatenation of a set of 3D points. 
$\mathbf{y}_0$ is the input pose and $\bar{\mathbf{y}}_0$ is root-shifted version of $\mathbf{y}_0$.  $\mathbf{P}$ is the diagonal matrix and its diagonal entries are either $1$ or $0$,   indicating whether the corresponding entry of the input pose is available or not. Through this introduction of $\mathbf{P}$,   we allow the input observation to be incomplete  while maintaining the formula integrity for complete observation by setting $\mathbf{P}$ to be the identity matrix.  This can conveniently model the users' constraints on specifying the fixed and moving (or missing) joints. 

The final term provides a prior constraint on the rotation parameters,   where the diagonal matrix $\mathbf{W}$ gives a weight for each of the 3 rotation parameters. Usually, the weight on the second rotation parameter $\tau_y$  (rotation around y-axis) shall be larger than the other two, as this rotation is usually more common.

By solving this problem, we find a pose that on one hand stays close to the input pose subject to a similarity transform, and on the other hand admits a sparse representation given the learned dictionary.  The input pose can be incomplete or corrupted by noise, and it can also consist of 2D point clouds obtained from an image, as showed in our experiments in next section.

The optimization variables in the problem $ (P_0)$ are the output pose $\mathbf{q}$, sparse coefficients $\mathbf{x}$, rotation parameters $\tau$ and positive scaling $s$. To solve the problem, we first find $\mathbf{x}$ and $\tau$ alternatively: in each iterate we first fix $\tau$ and find  $\mathbf{x}$ using OMP algorithm, and then fix $\mathbf{x}$ to find $\tau$ by gradient descend.  Based on the $\mathbf{x}$ and $\tau$ found, we then calculate the scaling $s$ by an algorithm referred to as Pose Normalization,   after which we finally determine the output pose Jacobian method. The framework for solving problem $ (P_0)$ is showed in figure \ref{fig_framework}, and the optimization details are presented in the next subsection. 

\subsection{Solving the pose synthesis problem $ (P_0)$}

To efficiently solve the problem $ (P_0)$,   we first assume that given $\mathbf{x}$,   we can find a positive $s$ and a $\mathbf{q}$ such that $\mathbf{Ax}=s \cdot f (\mathbf{q})$  (approximately) holds. By substituting it to \eqref{eqn_overall_objective},   we arrive at the following,  
\begin{eqnarray}
 (P_1) &\min. &||\mathbf{P}\{\tau\circ  (\mathbf{Ax}) - \bar{\mathbf{y}}_0\}||_2^2+||\mathbf{W} (\tau-\tau_0)||_2^2 \label{eqn_obj_p1}\\
&\text{s. t. }& ||\mathbf{x}||_0 \le \kappa
\end{eqnarray}

We use the alternating minimization framework to solve problem $ (P_1)$. More specifically,   we first solve the sparse coding problem by fixing $\tau$,    
\begin{eqnarray}
&\min. &||\mathbf{P}\{\mathbf{Ax} - \tau^{-1}\circ \bar{\mathbf{y}}_0\}||_2^2\\
&\text{s. t. }& ||\mathbf{x}||_0 \le \kappa, 
\end{eqnarray}
where $\tau^{-1}$ denotes the inverse of rigid-body rotation $\tau$. Let $\mathbf{v} = \tau^{-1}\circ \bar{\mathbf{y}}_0$,   then the above problem is equivalent to solving
\begin{eqnarray}
 (P_{1a})&\min. &||\mathbf{A}_p\mathbf{x} - \mathbf{v}_p||_2^2 \label{eqn_sparse_reduced} \\
&\text{s. t. }& ||\mathbf{x}||_0 \le \kappa \nonumber 
\end{eqnarray}
where $\mathbf{A}_p$ is extraction of rows of $\mathbf{A}$  which correspond to the non-zeros diagonal entries of $\mathbf{P}$,   and the same goes for $\mathbf{v}_p$. The problem $ (P_{1a})$ is solved by OMP.

We then find the rotation parameters $\tau$ by fixing the sparse coefficient $\mathbf{x}$. That is,   we solve the following unconstrained sub-problem: 
\begin{eqnarray}
 (P_{1b})&\min. &||\mathbf{P}\{\tau\circ  (\mathbf{Ax}) - \bar{\mathbf{y}}_0\}||_2^2+||\mathbf{W} (\tau-\tau_0)||_2^2 
\end{eqnarray}

The gradient information of $\tau$ can be used to solved this problem. Note again that the notation $\tau\circ \mathbf{y}$ denotes the operation that subsequently rotates the pose by three rotation angles around $x,  y$ and $z$ axis.  let $\mathbf{t}_1 =  (\frac{\pi}{2},  0,  0)$,  $\mathbf{t}_2 =  (0,  \frac{\pi}{2},  0)$,  $\mathbf{t}_3 =  (0,  0,  \frac{\pi}{2})$,   then the gradient $\mathbf{g}$ of the above objective function is given by
\begin{equation}
\mathbf{g}^{ (i)} = <\mathbf{P}\{\tau\circ \mathbf{y} - \bar{\mathbf{y}}_0\},  \mathbf{P}\{ (\tau+\mathbf{t}_i)\circ \mathbf{y}\}> + \mathbf{W}^{ (ii)} (\tau^{ (i)}-\tau_0^{ (i)})
\end{equation}
where $<,  >$ denotes inner product.

We alternatively solve the above two sub-problems $ (P_{1a})$ and $ (P_{1b})$ until convergence is reached. Once we have found the final sparse coefficients $\mathbf{x}$ and rotation parameters $\tau$,   we can determine the joint angles denoted as  vector $\mathbf{q}$ by solving the IK problem:
\begin{eqnarray}
 (P_2) &\min. &  ||\mathbf{Ax}-s\cdot f (\mathbf{q})||_2^2\\
&\text{s. t. }& s>0 \nonumber
\end{eqnarray}

Because of the involvement of Jacobian matrix, the Hessian for problem $ (P_2)$ is difficult to find. Moreover, Jacobian method,   gradient descend or quasi-Newton methods seem to be less efficient for this problem because of the unknown arbitrary scaling $s$.  Since we already know the lengths of all bones in our case,   we can leverage this  knowledge to determine the normalized pose $\tilde{\mathbf{y}}\overset{\underset{\mathrm{def}}{}}{=}\frac{1}{s}\mathbf{Ax}$. This process is referred to as Pose Normalization and is presented in algorithm \ref{alg_pose_normalization}. Note the normalization scheme is not simply making the pose vector normalized in $\ell_2$-norm sense.

\begin{algorithm}  

\caption{Pose normalization}          
\label{alg_pose_normalization}  
\begin{algorithmic}

\STATE \textbf{Input}: A pose $\mathbf{y}$ in Euclidean space,   standard bone length matrix $\mathbf{L}$. 
\STATE \textbf{Output}: A new pose $\tilde{\mathbf{y}}$
\STATE Divide the skeleton into five chains (see figure \ref{fig_pose_normalization}) as follows:
\begin{eqnarray}
&&\mathbf{c}_1 =  (1,  2,  3,  4,  5)\nonumber\\
&&\mathbf{c}_2 =  (1,  6,  7,  8,  9)\nonumber\\
&&\mathbf{c}_3 =  (1,  10,  11,  12,  13,  14,  15)\nonumber\\
&&\mathbf{c}_4 =  (12,  16,  17,  18,  19)\nonumber\\
&&\mathbf{c}_5 =  (12,  20,  21,  22,  23)\nonumber
\end{eqnarray}
\STATE Initialize $\Delta\mathbf{y}^{ (k)} = 0 $ for $k=1,. . . ,23$ and $\tilde{\mathbf{y}}^{ (1)} = \mathbf{y}^{ (1)}$
\FOR{$i = 1 \to 5$}
\FOR{ $j=2 \to$ length ($\mathbf{c}_i$)}
\STATE Let $k = \mathbf{c}_i[j]$,   $k-1 = \mathbf{c}_i[j-1]$, apply: 
\begin{eqnarray}
&&\tilde{\mathbf{y}}^{ (k)} = \Delta\mathbf{y}^{ (k-1)} + \frac{\mathbf{y}^{ (k)}-\mathbf{y}^{ (k-1)}}{||\mathbf{y}^{ (k)}-\mathbf{y}^{ (k-1)}||_2}\mathbf{L}_{k,  k-1} \nonumber\\
&&\Delta\mathbf{y}^{ (k)} = \tilde{\mathbf{y}}^{ (k)}-\mathbf{y}^{ (k)}\nonumber
\end{eqnarray}

\ENDFOR
\ENDFOR

where $\mathbf{y}^{ (k)}$ is the 3D coordinate of joint $k$. Joint $k-1$ is the parent of the joint $k$,   as already arranged so in the chain.  $\mathbf{L}_{k,  k-1}$ is the standard bone length between joint $k$ and $k-1$. $\tilde{\mathbf{y}}^{ (k)}$ is the new position of joint $k$.  

\end{algorithmic}
\end{algorithm}

\begin{figure}
\centering
\includegraphics[width = 0.8\linewidth]{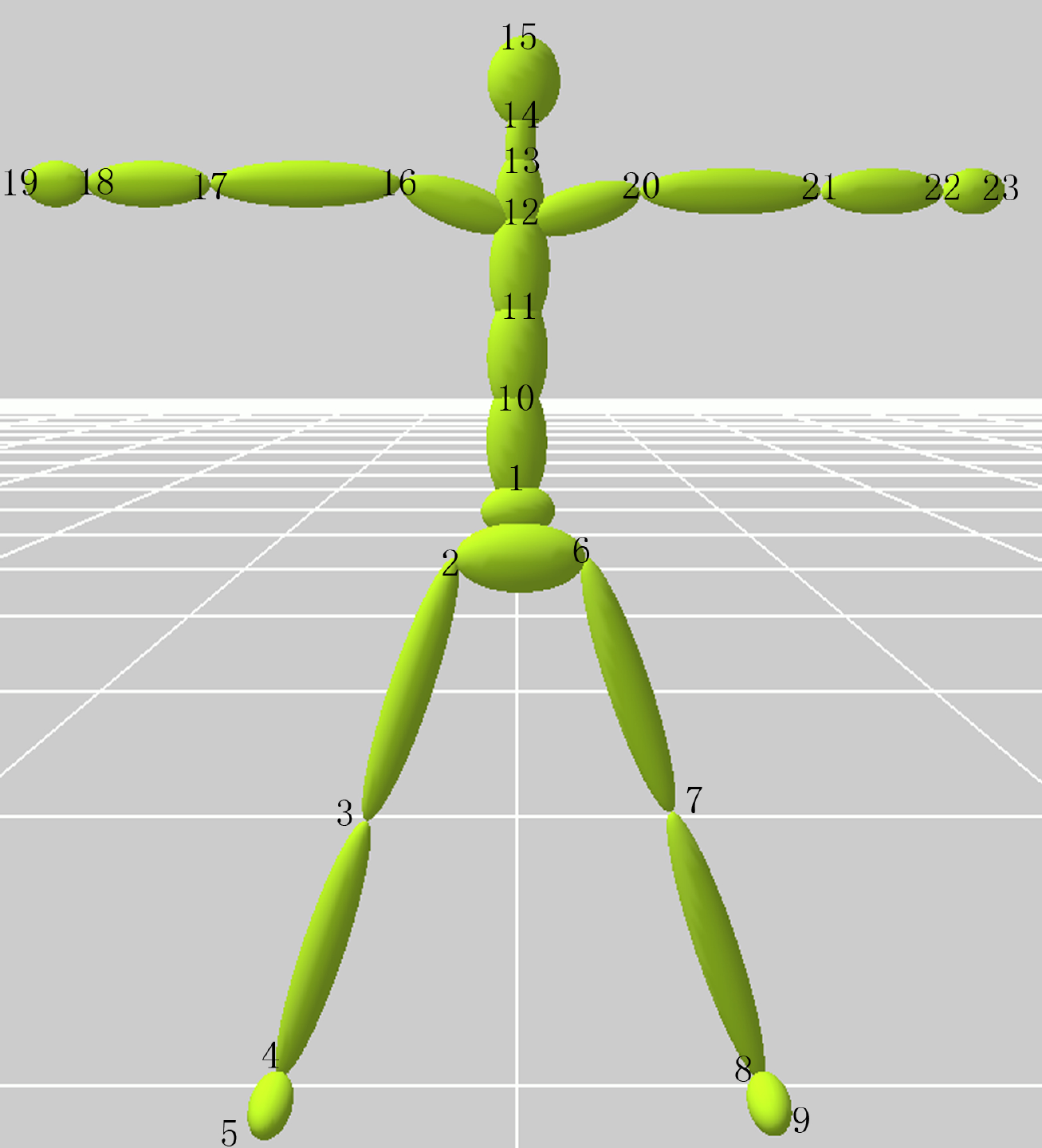}
\caption{Skeleton configuration. Each number is a joint ID labelling the corresponding joint.The skeleton can be divided into five chains: the torso, both arms and both legs.}
\label{fig_pose_normalization}
\end{figure}

After finding the normalized pose $\tilde{\mathbf{y}}$,   it is ready to apply Jacobian method  \cite{girard1985computational} to find $\mathbf{q}$ by solving the following non-linear system
 \begin{eqnarray}
f (\mathbf{q}) = \tilde{\mathbf{y}}
\end{eqnarray}
\textbf{Convergence and complexity}
By breaking the pose synthesis problem $ (P_0$) in to sub-problems $ (P_1)$ and $ (P_2)$, we greatly reduce the problem complexity. The assumption for this break-down is that the residual of the term in \eqref{eqn_overall_objective} diminishes, which corresponds to setting $\gamma$ to a large value. Thus this assumption holds and the break-down makes sense. The convergence of problem $ (P_1)$ is guaranteed as in each iterate both OMP and gradient descend decrease the objective value and the objective \eqref{eqn_obj_p1} is bounded below. 
The convergence of $ (P_2)$ is also guaranteed as the pose normalization algorithm is deterministic with constant complexity and the Jacobian method with complete and normalized target pose usually converge within $20$ iterates in our experiments.

\section{Applications and Experiments}

\subsection{Experimental setting}
The training samples we used are obtained from CMU motion capture website\footnote{http://mocap. cs. cmu. edu:8080/allasfamc. zip}. We manually trim the toes and fingers from the pose data.  These data are originally in 62 dimensional angle space, and they are trimmed to 46 dimensional so that the resulting skeletons contain only significant DOF's,  as in  \cite{wei2011intuitive}. When converted to Euclidean space, the skeleton model has $23$ joints with each in three dimensional . The total dimension for a pose is $D=69$. We pre-shift all the training samples to be rooted at the origin and set the global orientation to zeros. The corresponds to setting the first six components in the angle vector to zeros.

 We determine the size of pose dictionary $n$ by the following procedure: given a target learning error $e_t$, we randomly sample $n_0$ pose for the pose dictionary and use it to test the sparse coding error $e_s$ of the training set. If $e_s \le \delta e_t$, set $n$ to $n_0$ and use the current sampled poses as initialization for dictionary learning algorithm; otherwise, set $n_0$ to $1. 5n_0$ and continue the searching. We use $e_s\le \delta e_t$ as a criterion with $\delta$ usually set to $2$ because the dictionary learning can usually decrease the error by $50\%$, as we found in our experiments.

\subsection{Large-scale Comparison}

In large-scale comparison, we use the whole database from CMU, which sums up to 4150384 poses,   covering a varsity of motion styles ranging from basic types such as walking,   running to more complex types such as basketball and  golf. We randomly sample $50\%$ of poses for training all models and the rest for testing.  In training,   we set the cardinality upper-bound $\kappa$ to $3$.  The resulting pose dictionary consists of $262759$ atoms.

We test our model and other models using the testing set which consists of 2075280 poses. These models are MFA, the model with a Gaussian prior, the model with clustering and then build a Gaussian prior for each cluster (CG), LPCA and PCA.  The testing scheme is as follows. 

As we mentioned,   the existing models can be generalized to \eqref{eqn_general_data_driven}  (except for PCA which will be discussed  soon),   in which the parameter $\lambda$ is important in determining the resulting pose,   and its value provides a trade-off between the prior and the likelihood.  If the noise level is high,   the prior should be trusted more than the likelihood,   thus $\lambda$ should be decreased.  The same goes for $\kappa$ in our model.  Therefore,   the value of $\lambda$ for all other models and $\kappa$ for our synthesis model are chosen using brute-force search for a fair comparison.  Specifically,   to find the approximately best value,   we first randomly sample $0.1\%$ poses (about 2000) poses from the training set and use them to select the best value within a proper interval.  Then the best value for each model is used for testing the whole testing set.  

For MFA,   we use the same setting as stated in the paper \cite{wei2011intuitive},   except for the $\lambda$,   which was not given originally and is found by brute-force search.  we set the cluster number to $20$ for CG,   neighbourhood size to $100$ for LPCA.   For PCA, we use the first several principal components  such that $90\%$ energy of the corresponding eigenvalues is preserved.   Since the SIK is too computation-demanding,   we have omitted it from this large-scaled comparison.

\begin{figure}
\centering
\includegraphics[scale=0.2]{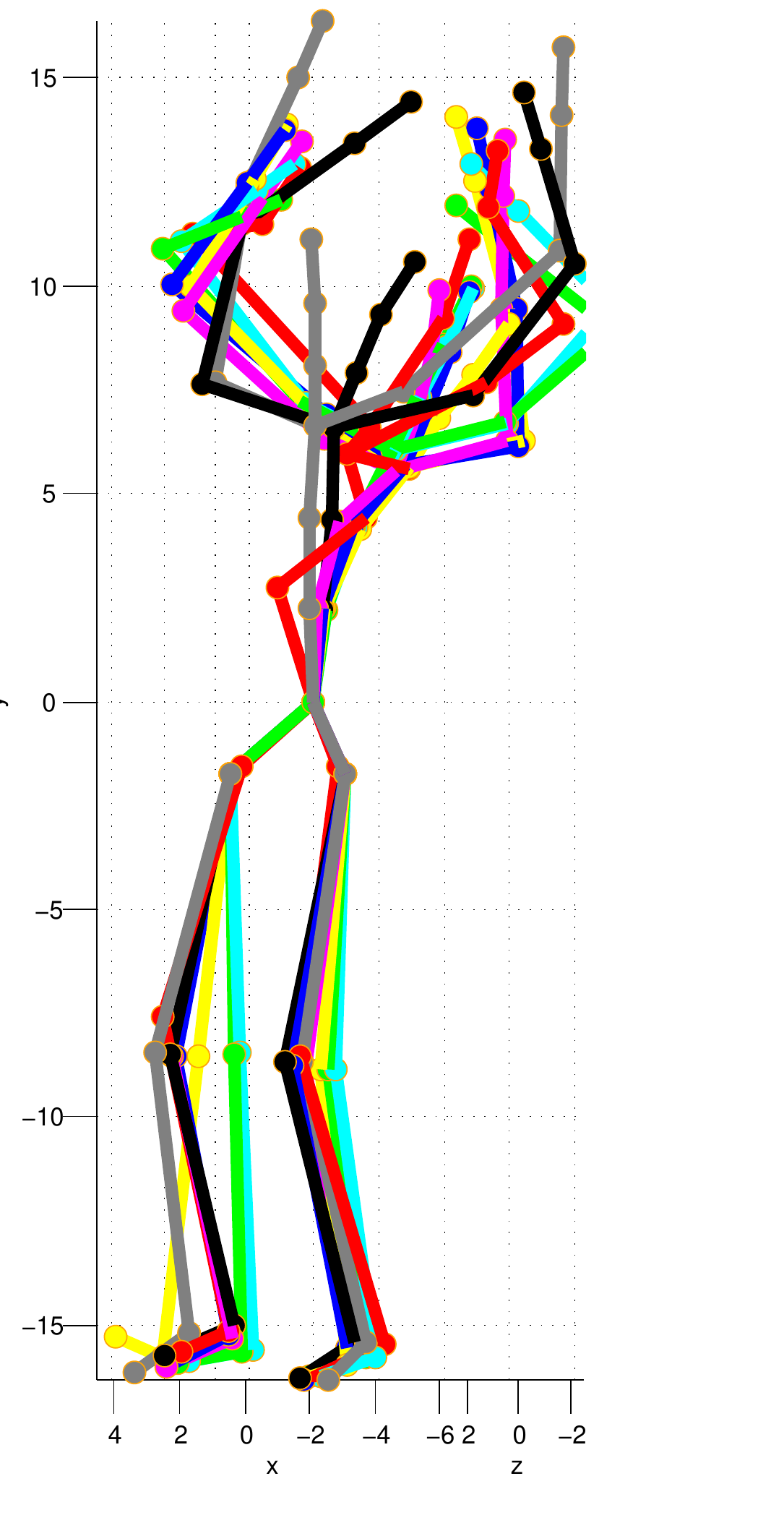}
\includegraphics[scale=0.2]{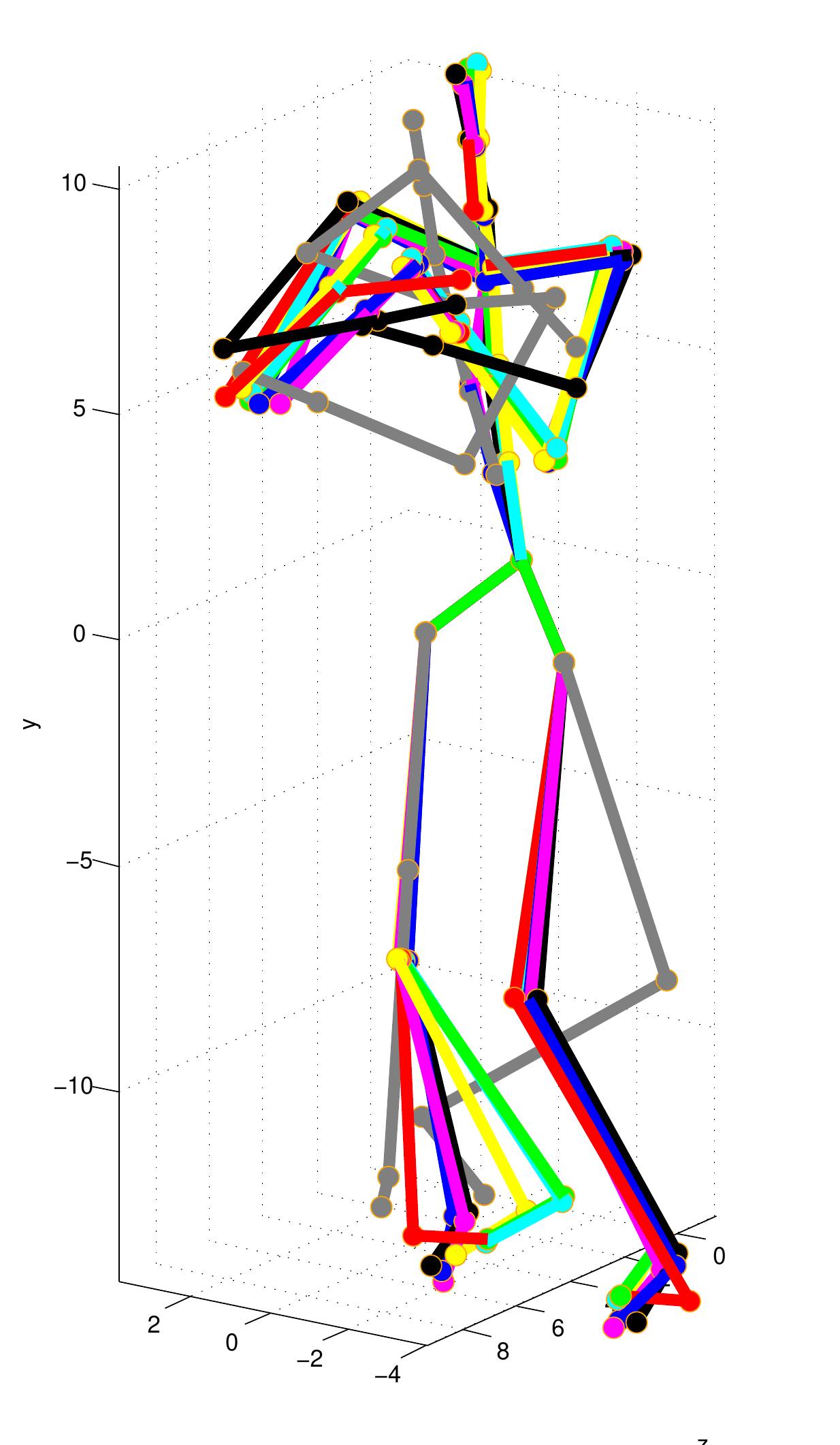}
\includegraphics[scale=0.2]{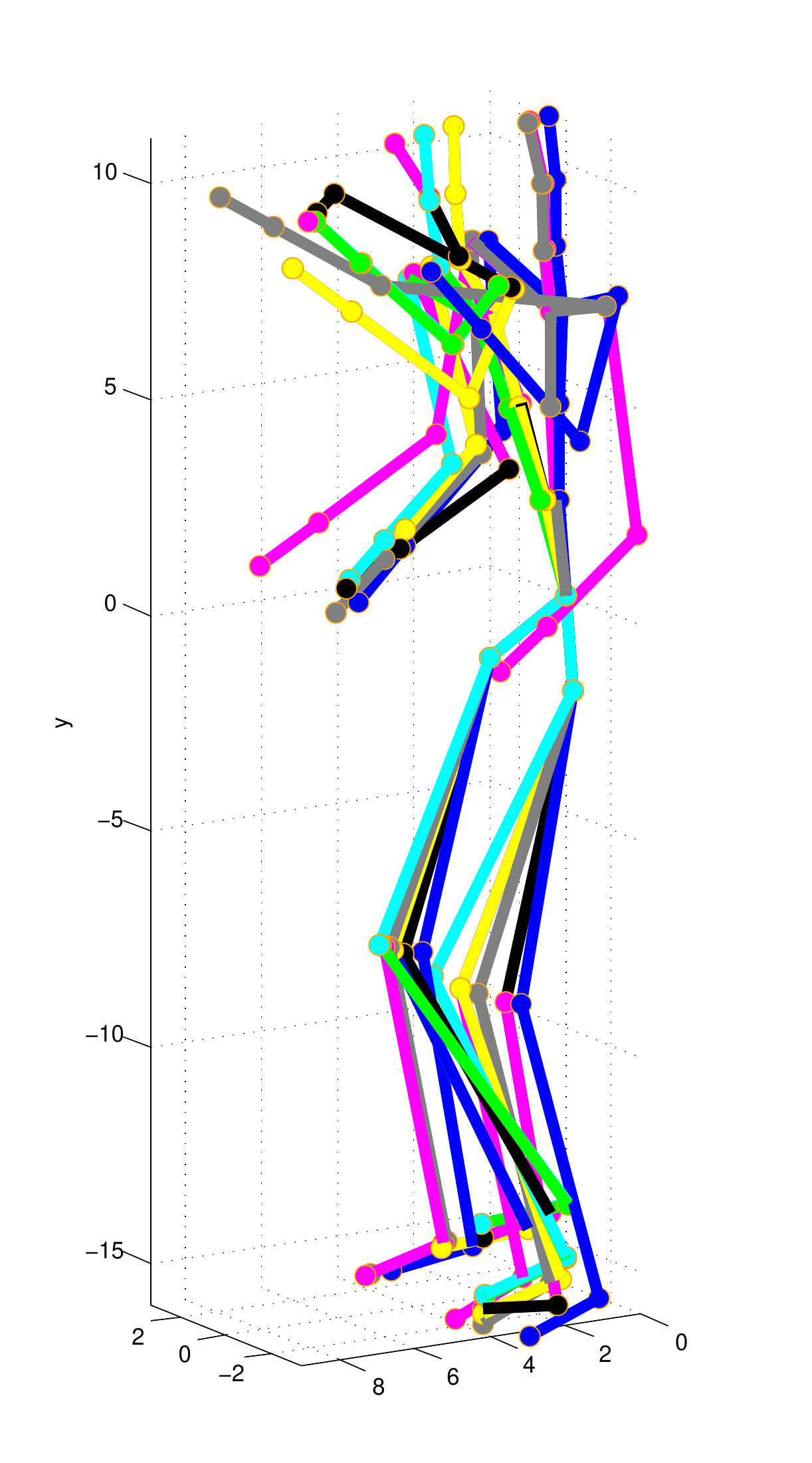}
\caption{A visual snapshot of the large-scale comparison. From left to right:  de-noising result when the noise is dense, sparse and completion result when only five joints (ID: 16,  20,  19,  23,  5 and 9) are observed. green: ground-truth, red: corrupted, cyan: our model, black: MFA, magenta: Gaussian, yellow: LPCA, blue: CG, gray: PCA}
\label{fig_visual_compare_three_tasks}
\end{figure}

We test the performance of de-noising and pose completion.  For de-noising,   we test two types of noise: dense and sparse Gaussian noise.   For dense Gaussian noise,   we generate standard Gaussian noise and add to the testing set.   For sparse Gaussian noise,   we generate standard Gaussian noise and randomly corrupt $20\%$ of the joints.   The mean square error (MSE) for each recovered pose is calculated and the average MSE for the whole testing set is showed in figure \ref{fig_large_compare}.  We also test the performance the completion when only a small portion of joints are observed.  Specially,   the inputs are the 3D coordinates of  joint ID 16,  20,  19,  23,  5 and 9 (see figure \ref{fig_pose_normalization} for joint ID map).  This missing pattern is the same as that in  \cite{wei2011intuitive}.   The comparison result is  showed in  \ref{fig_large_compare}.   For a visual instance of the large-scale comprison, see figure \ref{fig_visual_compare_three_tasks}.  

\begin{table}

\resizebox{7.5cm}{!} 
{
\centering
\begin{tabular}{|c|c|c|c|c|c|c|}
\hline 
Task&Our model & MFA & Gaussian & LPCA & CG & PCA\\ 
\hline 
\hline
dense noise&0.12 &  0.26 & 0.25 & 0.20 & 0.95 & 3.34\\ 
\hline
Sparse noise&  0.05 &  0.15 & 0.14 & 1.02 &  0.17 & 1.36 \\ 
\hline 
completion &0.01 &  0.24 & 0.30 & 0.80 & 0.20 &4.11\\ 
\hline 
\end{tabular} 
}
\caption{Average MSE of Large-scale comparison for de-noising and completion. The total testing poses for all the tasks are 2075280. As we see, our model performs better than other models for all the three tasks. PCA obtains considerably large errors for all tasks because the principal Components can not explain such a large dataset. }
\label{fig_large_compare}
\end{table}

\begin{table}[!]
\centering
\resizebox{7.5cm}{!} 
{
\begin{tabular}{|c|c|c|c|c|c|c|c|c|c|}
\hline 
Subject&No. Frames& Our model & MFA & Gaussian & LPCA & CG & SIK &PCA\\
\hline
\hline 
07& 2161& 0.02 &  0.13 & 0.08 & 0.07 & 0.02 & 0.70 & 0.07\\ 
\hline 
09& 769&0.03 &  0.08 & 0.10 & 0.06 & 0.06 & 1. 26 & 0.09\\ 
\hline 
63& 7529&0.07 &  0.41 & 2. 44 & 0.35 & 0.34 & 3. 90 & 0.41\\ 
\hline 
102&4252& 0.10 &  1.39 & 0.17 & 0.08 &  0.11 & 1.37 &1.28\\ 
\hline 
\end{tabular} 
}
\caption{Average MSE of small-scale comparison for de-noising when the noise is dense standard Gaussian noise. For subject 07, which contains similar styles of walking motions,  all the models perform  very well.  However, when the motions become more complex, for example, subject 63, our model outperforms all other models.  }
\label{fig_small_compare_dense_noise}
\end{table}

\begin{table}[!]
\centering
\resizebox{7.5cm}{!} 
{
\begin{tabular}{|c|c|c|c|c|c|c|c|c|}
\hline 
Subject&No. Frames& Our model & MFA & Gaussian & LPCA & CG & SIK &PCA\\
\hline 
\hline
07& 2161& 0.03&	0.08&	0.05&	0.07&	0.19&	0.26 & 0.08\\ 
\hline 
09& 769&0.04&	0.05&	0.06&	0.06&	0.05&	0.70 & 0.05\\ 
\hline 
63& 7529& 0.03&	0.12&	0.47&	0.09&	0.10&	0.99 & 0.36\\ 
\hline 
102&4252&  0.05&	0.08&	0.09&	0.07&	0.08&	1.34 &1. 09\\ 
\hline 
\end{tabular} 
}
\caption{Average MSE of small-scale comparison for de-noising when the noise is sparse standard Gaussian noise. In this test, MFA, LPCA and our model are all robust to change of dataset size and motion styles, and our model performs slightly better among the tree. }
\label{fig_small_compare_sparse_noise}
\end{table}

\begin{table}[!]
\centering

\resizebox{7.5cm}{!} 
{
\begin{tabular}{|c|c|c|c|c|c|c|c|c|}
\hline
Subject&No. Frames& Our model & MFA & Gaussian & LPCA & CG & SIK &PCA\\
\hline 
\hline
07&2161&0.07 &  0.11 & 0.23 & 0.25 & 0.95 & 0.31 & 0.07\\ 
\hline 
09&769&0.09 &  0.09 & 0.26 & 0.26 & 0.25 & 0.74 &0.05 \\ 
\hline 
63&7529& 0.07 &  0.18 & 0.26 & 0.18 & 0.21 & 3. 90 &  0.53\\ 
\hline 
102&4252& 0.09 &  0.24 & 0.28 & 0.25 &  0.27 & 1.48& 1. 30 \\ 
\hline 
\end{tabular} 
}
\caption{Average MSE of small-scale comparison for completion. In this test, our model is the only one that has obtained MSE smaller than 0.1 for all four subjects. }
\label{fig_small_compare_missing_infer}
\end{table}

\begin{figure}[!]
\centering
\includegraphics[width = \linewidth]{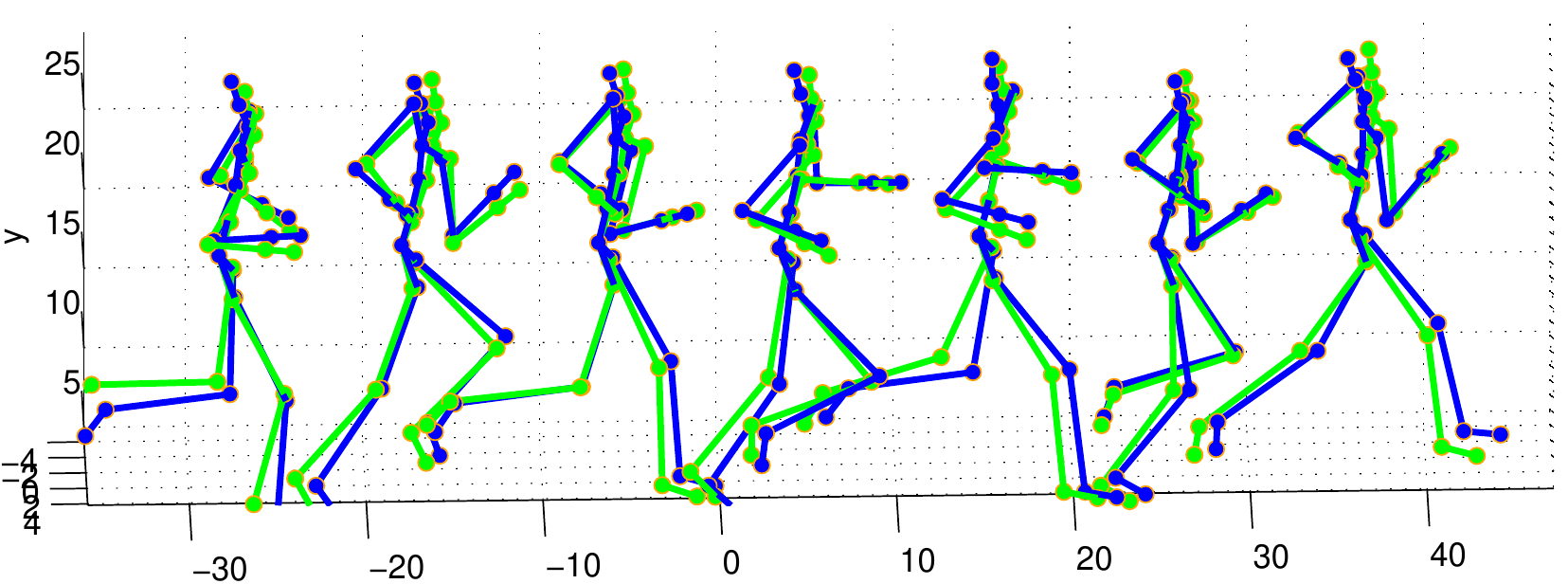}
\includegraphics[width = \linewidth]{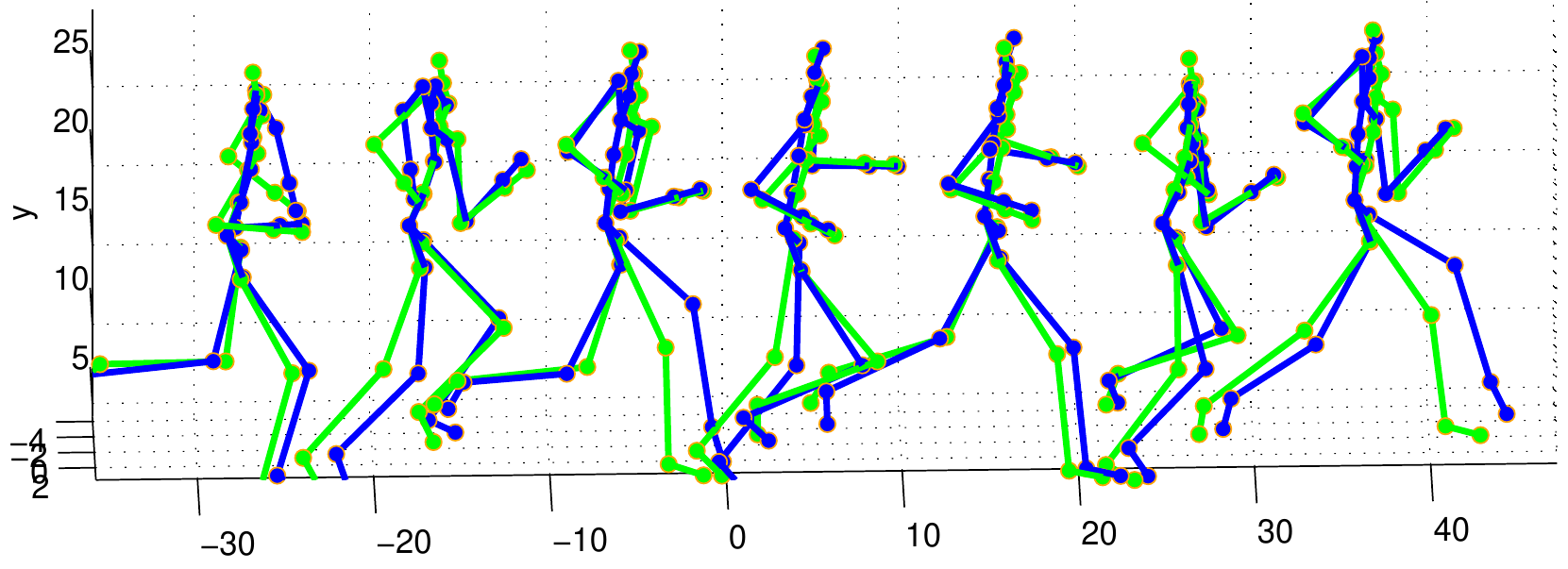}
\includegraphics[width = \linewidth]{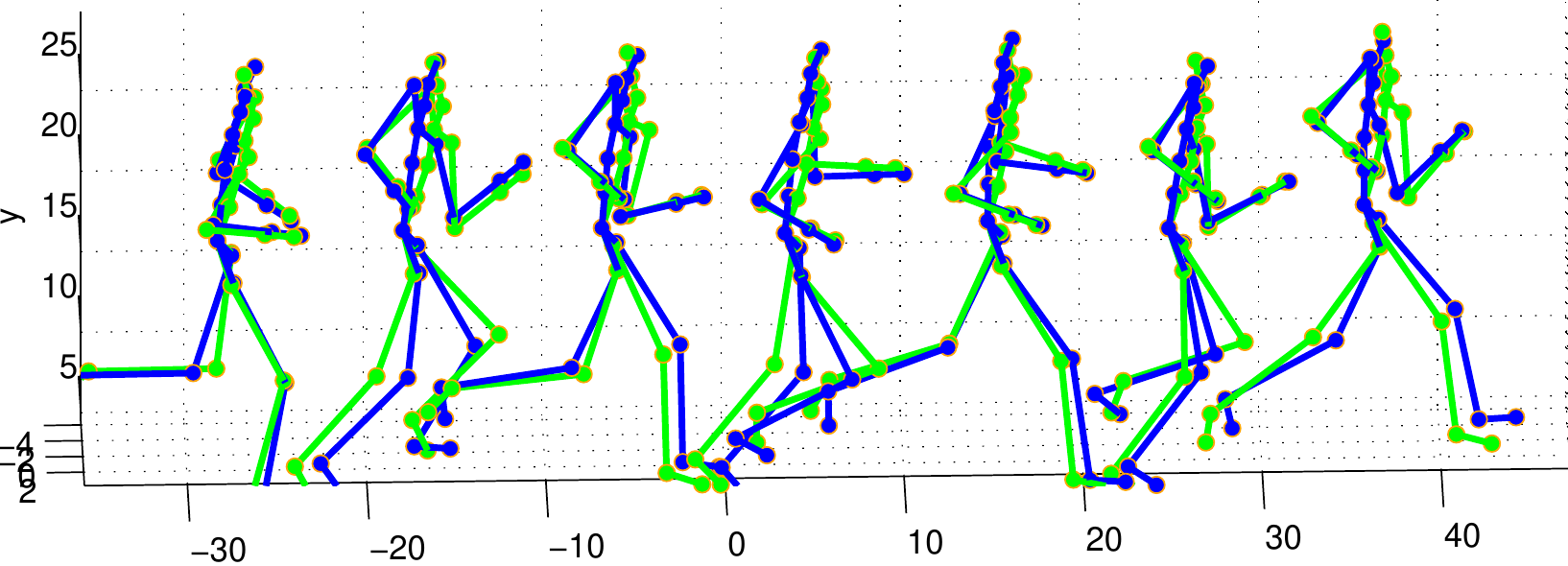}
\includegraphics[width = \linewidth]{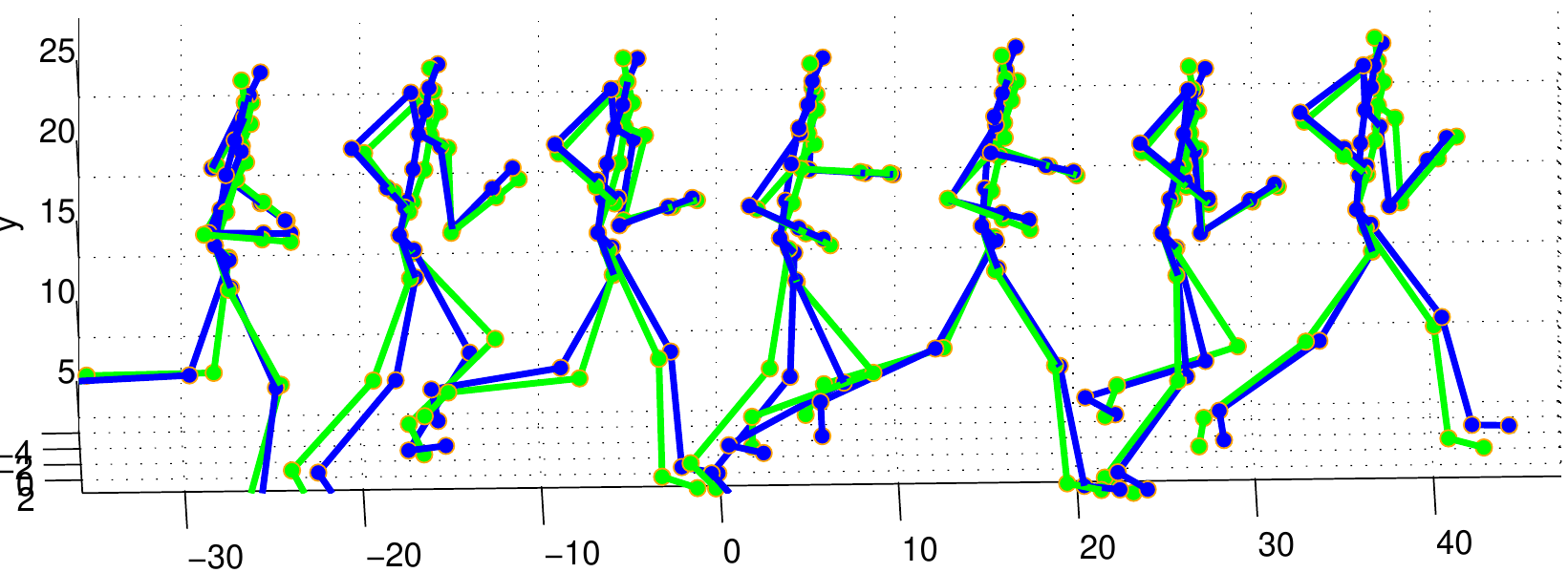}
\includegraphics[width = \linewidth]{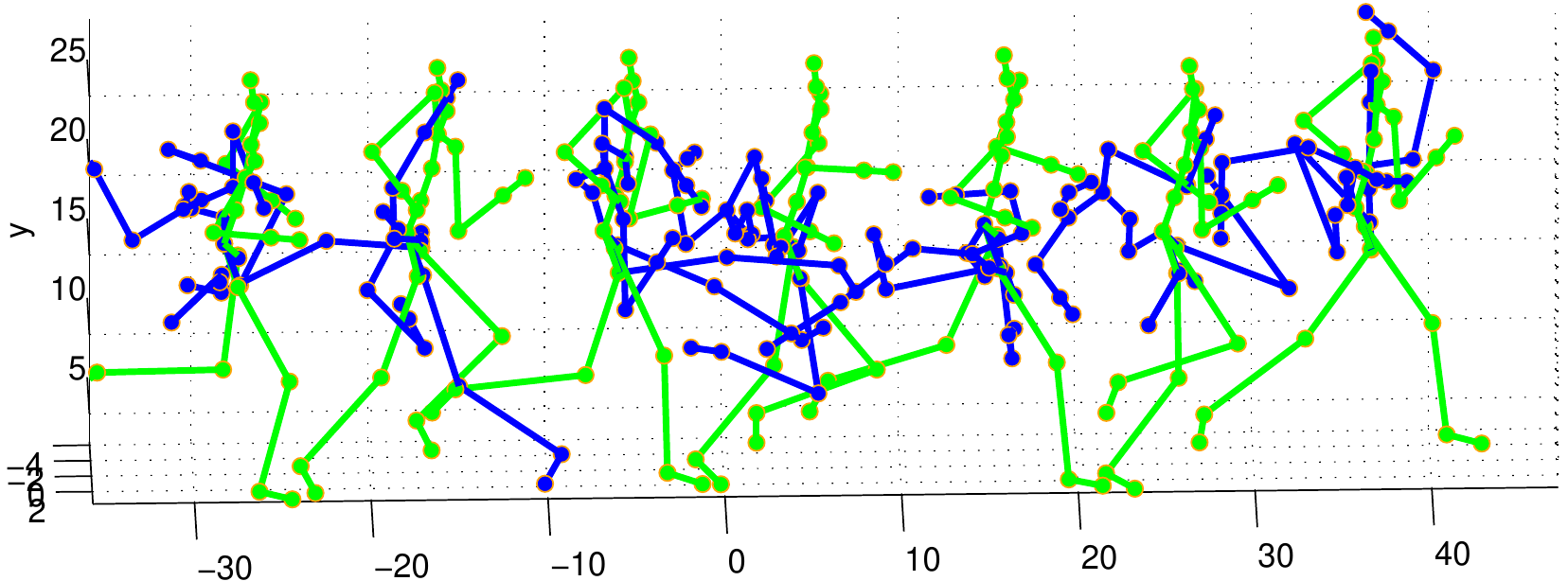}
\includegraphics[width = \linewidth]{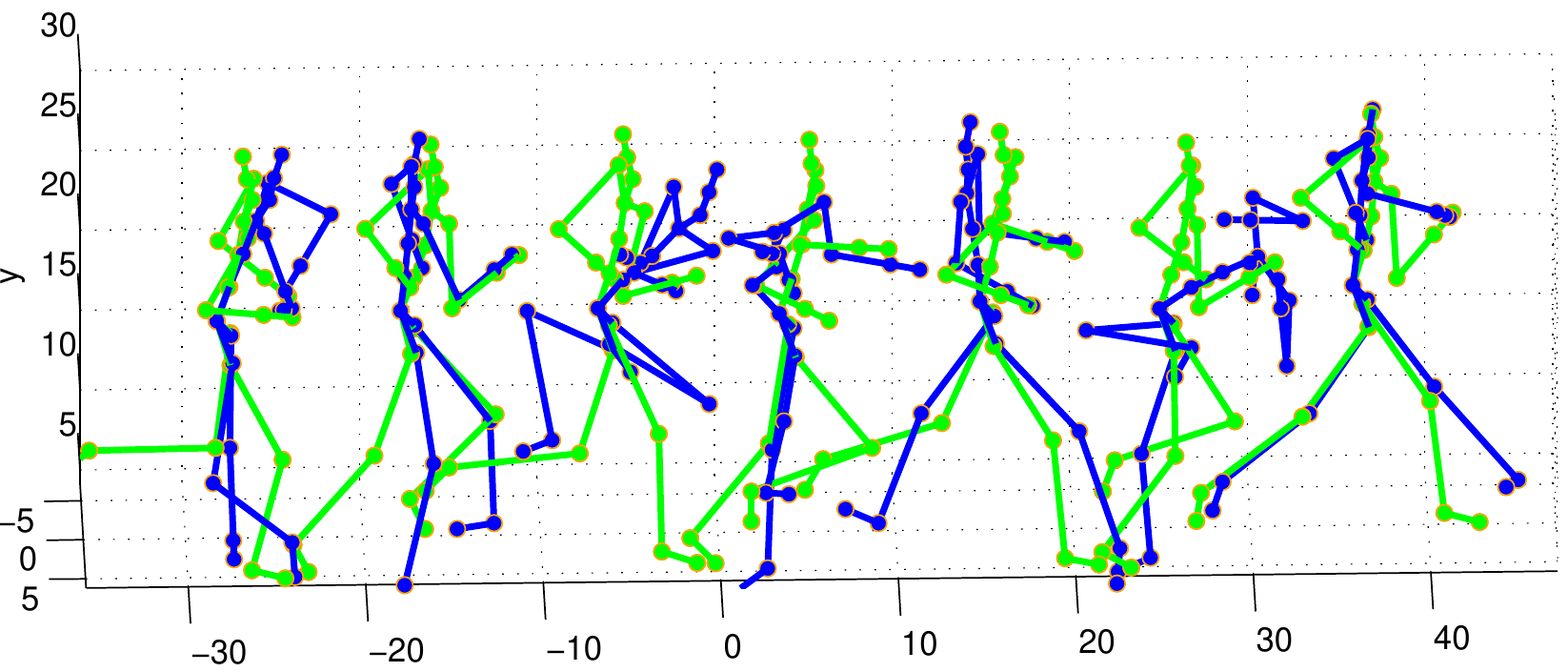}

\caption{Comparison of all models for completing the first 128 poses of a running sequence in subject 09 of CMU database. All models are trained as in the large-scale comparison. For a clear view, the  poses are showed at every 20 frames. The ground-truth is showed in green together with the results obtained from all models (in blue) which are respectively from top to bottom: our model, MFA, Gaussian, CG, LPCA and PCA. Our model performs best as it preserves a better pose structure. For the rest models, Gaussian, CG  and MFA fail to capture the running style of the upper-body (from back-bone to head) and defects can also be spotted in the head and feet. Neither LPCA nor PCA obtains an acceptable result in this comparison. }
\label{fig_motion_completion}
\end{figure}

Even though all models considered in this paper do not take into account the motion dynamics  in training stage, we can still compare their performance on motion completion by applying completion algorithm to each  (incomplete) pose in the motion, as this provides a good reflection on the performance of pose complete.  This comparison is done to a running motion and the result is showed in figure \ref{fig_motion_completion}.  Our model outperforms other models in that it preserves a better pose structure of the upper-body (see the figure caption for more details).

\subsection{Small-scale Comparison}

Similar to the above large scale comparison,   we also test the performance of each model for learning from small datasets.  We choose four subjects from CMU mocap database website: 07 (walking),   09 (running),   63 (golf) and 102 (basketball). These subjects are representative as they are different styles of motion  and their size varies from 1538 to  15079 poses. For each subject,   we randomly sample $50\%$ for training and the rest for testing. The training and testing schemes are the same as that in large scale comparison. In training stage,   the setting for SIK is the same as mentioned in  \cite{grochow2004style}; for other models,   the setting is similar to the above. We also test the performance of completion and de-noising under two types of noises. The results for these three tasks are showed the table \ref{fig_small_compare_dense_noise},  \ref{fig_small_compare_sparse_noise} and \ref{fig_small_compare_missing_infer} respectively. 
 As we see,    our model outperforms the other models for three tasks even for small datasets.

\subsection{Interactive character posing}

We provide an real-time application of our model in interactive character posing.  The user interface is implemented in C++ and we use the pose dictionary learned in the large-scale comparison for pose synthesis.  We consider two kinds of input here. Free-dragging interface provides a  freely-edited complete pose as an input. Pose completion takes a set of 2D or 3D points as inputs and reconstructs the whole pose.  Other inputs are possible, as long as they can fit into the model, perhaps after some necessary preprocessing . 

\textbf{Free-dragging}
A common scenario is that when the user drags one or multiple joints of the skeleton,   the computer is required to respond to this drag and create a new pose. Chances are that the edited pose looks like being corrupted by noise if the user is novice. To synthesize a new pose based on the corrupted pose and the pose dictionary ,   $ (P_0)$ is  solved with $\mathbf{P}$ set to the identity matrix.  This corresponds to setting all joints of the input pose $\mathbf{y}_0$ as soft-constraints.  

\begin{figure}
\centering
\includegraphics[width = \linewidth]{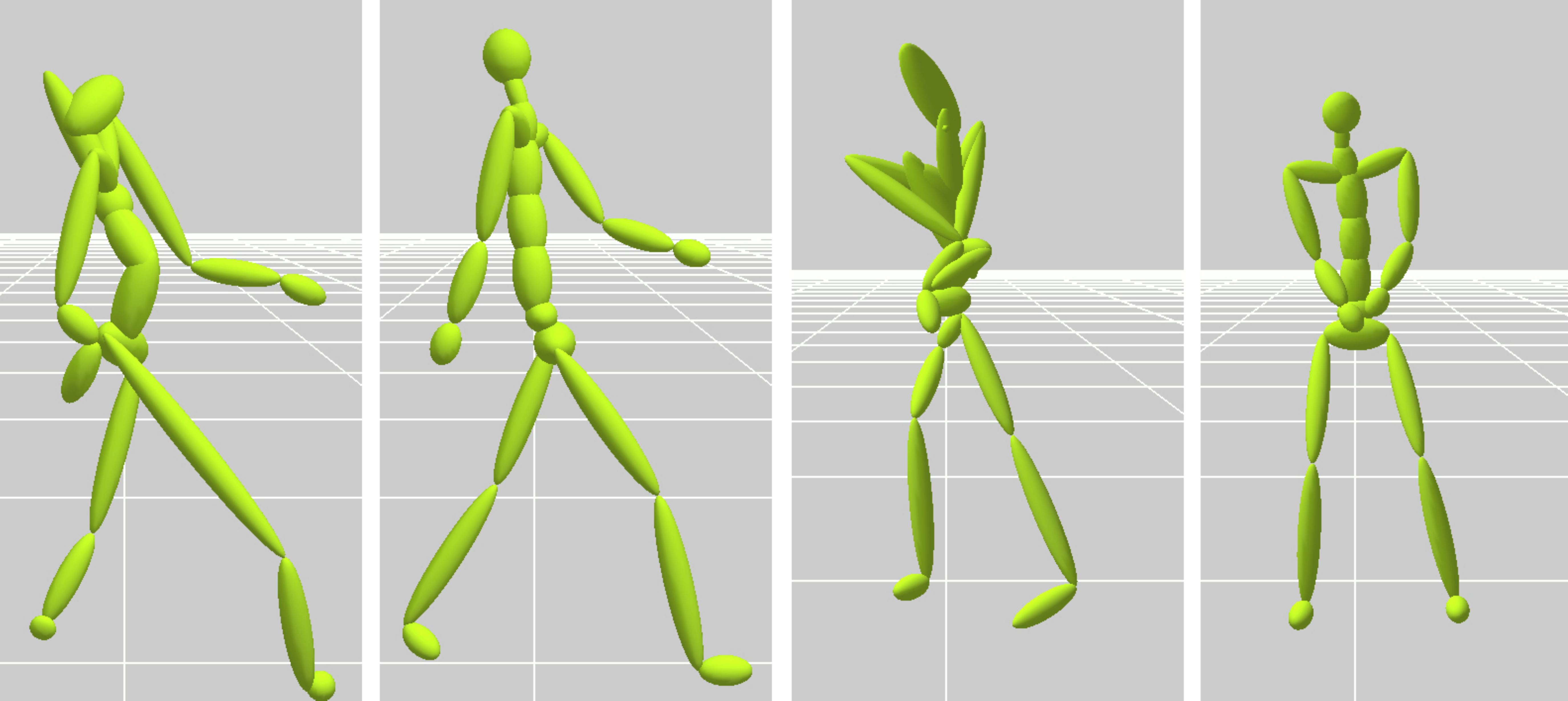}

\caption{Some examples of free-dragging and pose reconstruction results. First and third: modified poses; Second and fourth: the corresponding reconstructed poses }
\label{fig_UI_example}
\end{figure}

As our model is trained from the pose data in Euclidean space, it is well-suited for interactive character posing in which the user can arbitrarily modify the pose without being worried about bone-length constraints and angle limits. This provides a great continence for the user because s/he can now move the joints to wherever s/he wants. After the user finishes the modification, our model synthesizes natural poses that satisfies the user's intention on the style and corrects all the violations. See figure \ref{fig_UI_example} for examples. 

\begin{figure}[!]
\centering
\includegraphics[width = 0.45\linewidth]{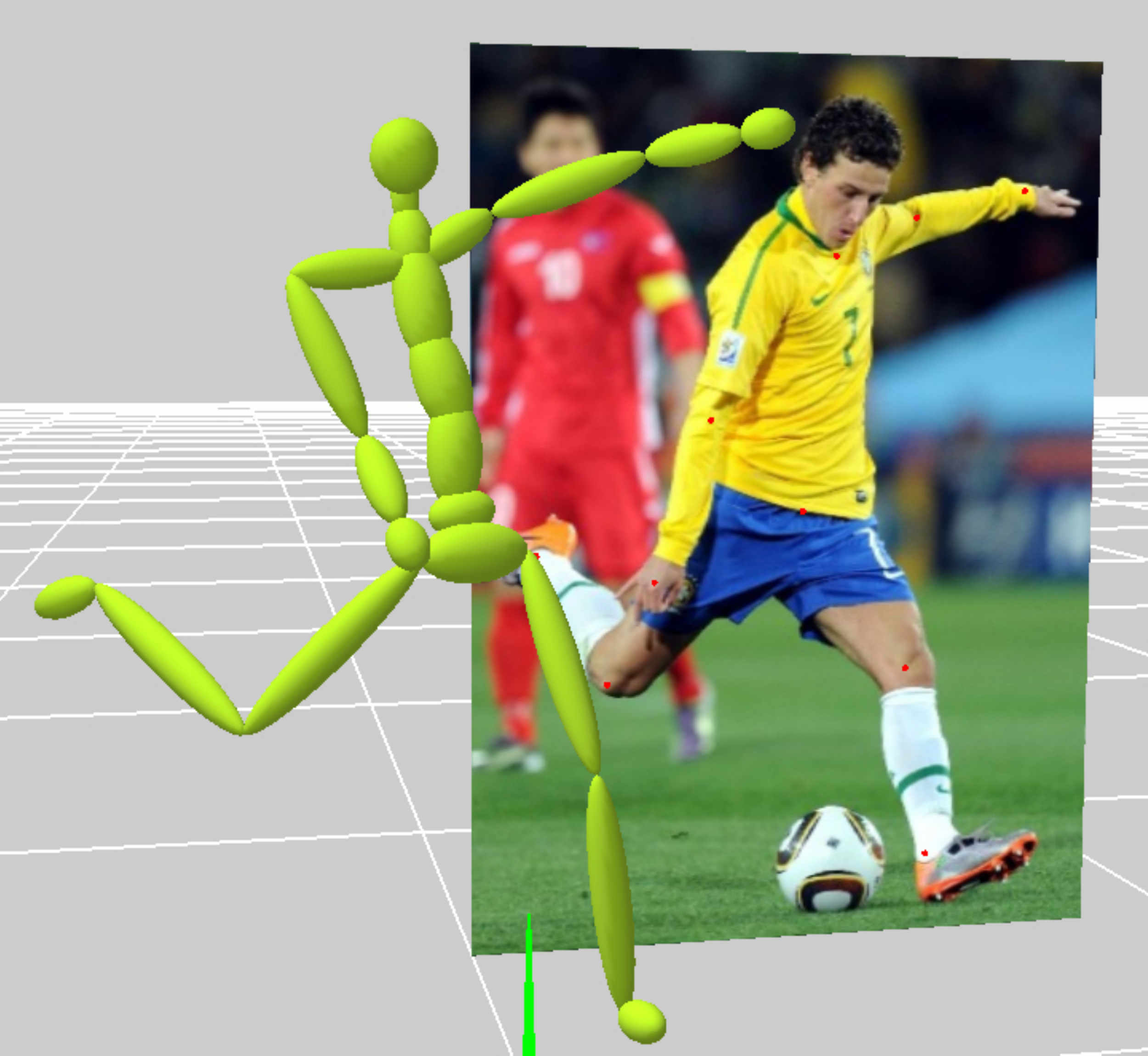}
\includegraphics[width = 0.40\linewidth]{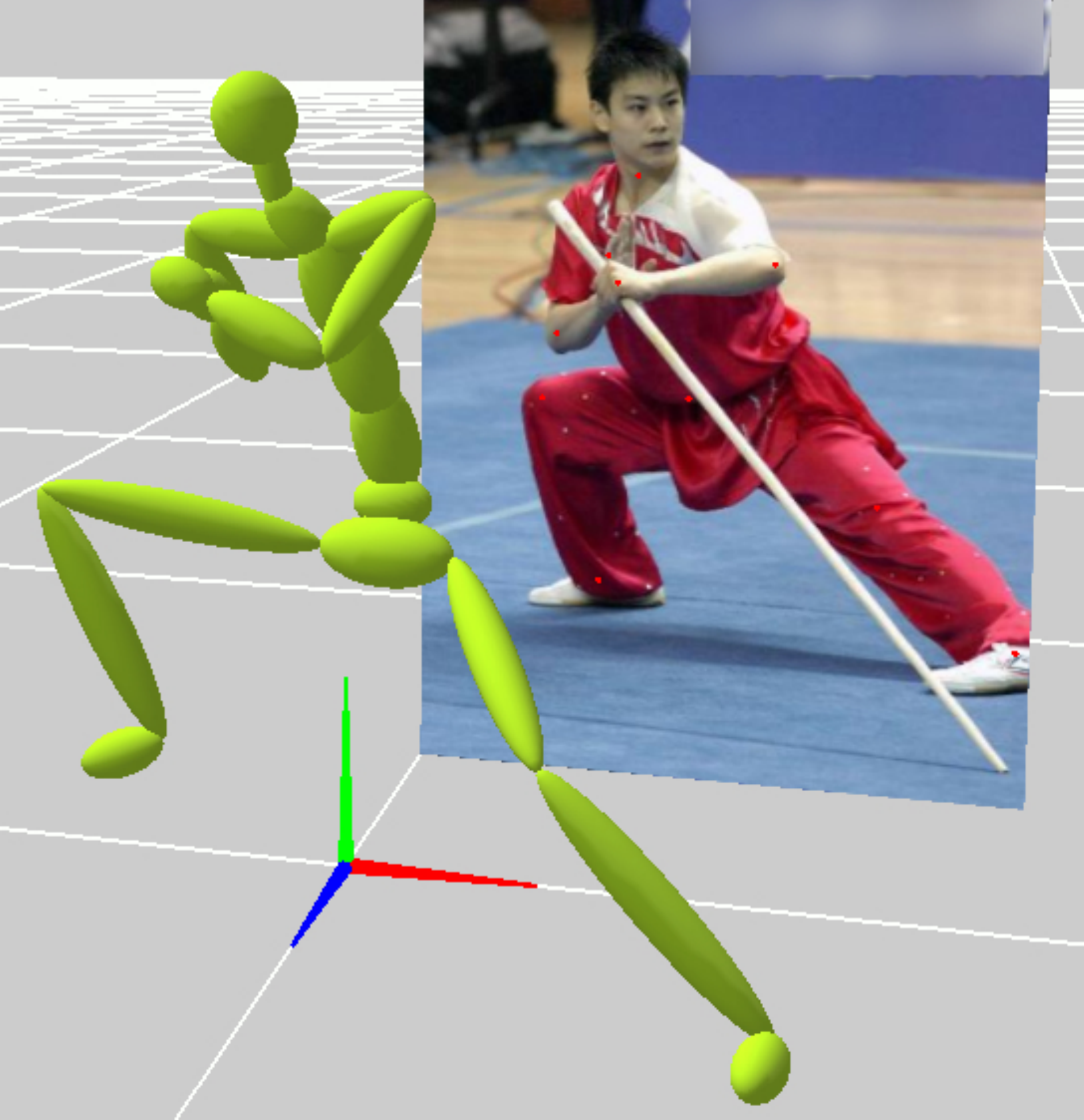}

\caption{Reconstructing 3D poses given 2D inputs showed as red dots. The 2D coordinates of the inputs are assigned to the corresponding joints(chosen by users) for pose reconstruction. }
\label{fig_d2D_reconstruct}
\end{figure}

\textbf{Pose completion}
In pose completion problem,   we infer the whole pose while only a portion of the joints are observed. It turns out that our model can be conveniently adapted to solving this problem even when the model is trained with the full-body pose data.  To do this,   we simply set the entries of $\mathbf{P}$ corresponding to the joints that we want to set as observed (fixed) to ones and the rest to zeros.  With this introduction of $\mathbf{P}$,   we can conveniently incorporate 2D inputs: given a  picture which contains a pose,   the user can label the joints with the 2D coordinate of the pose in the picture and reconstruct a 3D pose.  We give two examples in figure \ref{fig_d2D_reconstruct}.

\section{Discussions and conclusions}

\textbf{De-nosing and completion}
We have compared our model with the existing ones for the performance of de-noising and completion. One may think that de-noising is irrelevant as the motion capture data are usually 'clean'. This perhaps is true for the already-available databases. In the process of motion capture however, de-noising is necessary because of measurement error and  sensor failures\cite{rose1997process,lou2010example}.  Moreover, for interactive character posing, the pose edited by users is usually \textit{noisy} in the sense that it is inconsistent with the training set. We can see the process of pose editing as a measurement of users' intentions, which will always introduce measurement noise. Apart from dense noise, we have also considered sparse noise. This is meaningful because in motion capture process, noise can be sparse due to error introduced in a few sensors. Similarly, in character posing, the user may only edit a few joints, making the measurement noise sparse. Pose completion also is useful, not only in dealing with motion capture data when the measurements are incomplete, but also in character posing   to account for users' constraints.

\textbf{Sparsity}  The choice of $\kappa$ in pose synthesis stage depends on the noise level. It reflects our initial knowledge on the property of noise. If we believe that the noise level is high, we can reduce the $\kappa$; otherwise, we can increase $\kappa$ such that it can better approximate the input. This then offers a trade-off between stratifying  the users' exact constraints (which may result in an invalid pose) and synthesizing a realistic and natural pose.  

\textbf{Combining dictionaries} Our synthesis model is flexible in that it can combine dictionaries that are learned separately by simply Concatenating all sub-dictionaries.. This provides a friendly solution for  accepting large scale training set in the training stage.  In terms of complexity,   K-SVD is $O (m^2)$,   where $m$ is the size of training set. By using fast algorithms for clustering such as K-means,   we can divide the training set into smaller subsets with complexity $O (m)$ before applying K-SVD to each of them. In this way,   the overall training complexity is reduced. Since K-SVD is a generalization of k-means,   this approximation  is analogical to the  hierarchical version of k-means. We use this approach for the large scale training. 

\textbf{Physical constraints}
The only physical constraints we consider in this paper is the bone length. Angle limits are not considered here as we found that the solving the problem $ (P_0)$ usually will not violate the angle limit constraints.  However,   they can be interoperated into sub-problem $ (P_2)$ if necessary, and the problem can be solved for example similarly to  \cite{zhao1994inverse}. 

\textbf{Connection to subspace models}
By setting $\kappa$ to a large number and imposing extra orthogonal constraints on $\mathbf{A}$, the pose dictionary is the basis matrix obtained from subspace models and the pose synthesis problem $(P_0)$ is almost the same as the PCA model which optimizes the pose in PCA subspace(except that we model the pose in Euclidean space).  From this point of view, our model can be seen as a generalization of the  PCA subspace model.

\textbf{Connection to compressed sensing}
In compressed sensing\cite{donoho2006compressed}, the random sensing matrix plays an important role. Our model is related to compressed sensing except that we set the random sensing matrix to be  square. Then this sensing matrix will have no effect as we can take inverse and remove it from the model. That is, We do not perform any reduced measurement on the pose. This is because the input pose might be incomplete already, as indicated by $P$. Introducing a (fat) sensing matrix will complicate the (incomplete) measurement and make it more difficult to recover the pose.

\textbf{Connection to nearest-neighbour} 
Although similar in some sense,   our approach is not nearest-neighbour (NN) algorithm. Firstly,   our model is a parametric model,   while the NN algorithm is not. Secondly,   although the OMP algorithm used for sparse coding stage is a greedy algorithm that resembles NN,  it is in fact a greedy algorithm for the solving the sparse coding problem. Other algorithms such as linear programming ,   shrinkage and  interior point method can also be used.  However,   we find that OMP is more efficient in our case.

\textbf{Conclusion}
In this paper,   we have proposed a model for articulate character posing. We have shown that our model can be trained to  learn the pose dictionary from a large-scale training set. We also demonstrated how to apply our model in de-noising and completion problem. We have also provided UI examples showing how to use our model for character posing. Experiments have shown that our model outperforms the existing models in pose de-noising and completion.

One limitation of our model is that to achieve a small learning error, the pose dictionary size could be large for learning from a large dataset. This could be a problem for applications in devices that have limited memory. Nevertheless, our model is currently designed for applications in personal computers.

\section*{Acknowledgement}
This project is supported by the Faculty Research Grant of Department of Computer Science, Hong Kong Baptist University.

\bibliographystyle{eg-alpha}
\bibliography{template}

\newcommand{\etalchar}[1]{$^{#1}$}
\begin{thebibliography}{\uppercase{YWHM08}}

\bibitem[AEB06]{aharon2006ksvd}
\textsc{Aharon M., Elad M., Bruckstein A.}:
\newblock k-svd: An algorithm for designing overcomplete dictionaries for
  sparse representation.
\newblock \emph{Signal Processing, IEEE Transactions on 54}, 11 (nov. 2006),
  4311 --4322.

\bibitem[BH00]{brand2000style}
\textsc{Brand M., Hertzmann A.}:
\newblock Style machines.
\newblock In \emph{Proceedings of the 27th annual conference on Computer
  graphics and interactive techniques} (New York, NY, USA, 2000), SIGGRAPH '00,
  ACM Press/Addison-Wesley Publishing Co., pp.~183--192.

\bibitem[CH05]{chai2005performance}
\textsc{Chai J., Hodgins J.~K.}:
\newblock Performance animation from low-dimensional control signals.
\newblock \emph{ACM Trans. Graph. 24} (July 2005), 686--696.

\bibitem[Don06]{donoho2006compressed}
\textsc{Donoho D.}:
\newblock Compressed sensing.
\newblock \emph{Information Theory, IEEE Transactions on 52}, 4 (2006),
  1289--1306.

\bibitem[GH96]{ghahramani1996algorithm}
\textsc{Ghahramani Z., Hinton G.}:
\newblock The em algorithm for mixtures of factor analyzers.
\newblock \emph{University of Toronto Technical Report} (1996).

\bibitem[GM85]{girard1985computational}
\textsc{Girard M., Maciejewski A.}:
\newblock Computational modeling for the computer animation of legged figures.
\newblock \emph{ACM SIGGRAPH Computer Graphics 19}, 3 (1985), 263--270.

\bibitem[GMHP04]{grochow2004style}
\textsc{Grochow K., Martin S.~L., Hertzmann A., Popovi\'{c} Z.}:
\newblock Style-based inverse kinematics.
\newblock In \emph{ACM SIGGRAPH 2004 Papers} (New York, NY, USA, 2004),
  SIGGRAPH '04, ACM, pp.~522--531.

\bibitem[GMW81]{gill1981practical}
\textsc{Gill P., Murray W., Wright M.}:
\newblock Practical optimization.

\bibitem[Kal08]{kallmann2008analytical}
\textsc{Kallmann M.}:
\newblock Analytical inverse kinematics with body posture control.
\newblock \emph{Computer Animation and Virtual Worlds 19}, 2 (2008), 79--91.

\bibitem[Law04]{lawrence2004gaussian}
\textsc{Lawrence N.}:
\newblock Gaussian process latent variable models for visualization of high
  dimensional data.
\newblock \emph{Advances in neural information processing systems 16} (2004),
  329--336.

\bibitem[LC10]{lou2010example}
\textsc{Lou H., Chai J.}:
\newblock Example-based human motion denoising.
\newblock \emph{Visualization and Computer Graphics, IEEE Transactions on 16},
  5 (sept.-oct. 2010), 870 --879.

\bibitem[LFAJ10]{li2010learning}
\textsc{Li Y., Fermuller C., Aloimonos Y., Ji H.}:
\newblock Learning shift-invariant sparse representation of actions.
\newblock In \emph{Computer Vision and Pattern Recognition (CVPR), 2010 IEEE
  Conference on} (june 2010), pp.~2630 --2637.

\bibitem[LWS02]{li2002motion}
\textsc{Li Y., Wang T., Shum H.-Y.}:
\newblock Motion texture: a two-level statistical model for character motion
  synthesis.
\newblock In \emph{Proceedings of the 29th annual conference on Computer
  graphics and interactive techniques} (New York, NY, USA, 2002), SIGGRAPH '02,
  ACM, pp.~465--472.

\bibitem[LYL11]{lai2011motion}
\textsc{Lai R. Y.~Q., Yuen P.~C., Lee K. K.~W.}:
\newblock {Motion Capture Data Completion and Denoising by Singular Value
  Thresholding}.
\newblock Avis N., Lefebvre S., (Eds.), Eurographics Association, pp.~45--48.

\bibitem[PRK93]{pati1993orthogonal}
\textsc{Pati Y., Rezaiifar R., Krishnaprasad P.}:
\newblock Orthogonal matching pursuit: Recursive function approximation with
  applications to wavelet decomposition.
\newblock In \emph{Signals, Systems and Computers, 1993. 1993 Conference Record
  of The Twenty-Seventh Asilomar Conference on} (1993), IEEE, pp.~40--44.

\bibitem[RRP97]{rose1997process}
\textsc{Rose B., Rosenthal S., Pella J.}:
\newblock The process of motion capture: Dealing with the data.
\newblock In \emph{Computer Animation and Simulation} (1997), vol.~97.

\bibitem[SHP04]{safonova2004synthesizing}
\textsc{Safonova A., Hodgins J.~K., Pollard N.~S.}:
\newblock Synthesizing physically realistic human motion in low-dimensional,
  behavior-specific spaces.
\newblock In \emph{ACM SIGGRAPH 2004 Papers} (New York, NY, USA, 2004),
  SIGGRAPH '04, ACM, pp.~514--521.

\bibitem[WC11]{wei2011intuitive}
\textsc{Wei X., Chai J.}:
\newblock Intuitive interactive human-character posing with millions of example
  poses.
\newblock \emph{Computer Graphics and Applications, IEEE 31}, 4 (july-aug.
  2011), 78 --88.

\bibitem[WTR11]{wu2011natural}
\textsc{Wu X., Tournier M., Reveret L.}:
\newblock Natural character posing from a large motion database.
\newblock \emph{Computer Graphics and Applications, IEEE 31}, 3 (may-june
  2011), 69 --77.

\bibitem[WYG{\etalchar{*}}09]{wright2009robust}
\textsc{Wright J., Yang A., Ganesh A., Sastry S., Ma Y.}:
\newblock Robust face recognition via sparse representation.
\newblock \emph{Pattern Analysis and Machine Intelligence, IEEE Transactions on
  31}, 2 (feb. 2009), 210 --227.

\bibitem[YWHM08]{yang2008image}
\textsc{Yang J., Wright J., Huang T., Ma Y.}:
\newblock Image super-resolution as sparse representation of raw image patches.
\newblock In \emph{Computer Vision and Pattern Recognition, 2008. CVPR 2008.
  IEEE Conference on} (june 2008), pp.~1 --8.

\bibitem[ZB94]{zhao1994inverse}
\textsc{Zhao J., Badler N.~I.}:
\newblock Inverse kinematics positioning using nonlinear programming for highly
  articulated figures.
\newblock \emph{ACM Trans. Graph. 13} (October 1994), 313--336.

\end{thebibliography}

\end{document}